\documentclass[prb,aps,twocolumn,nobibnotes,epsf]{revtex4}

\usepackage{graphicx}
\usepackage{dcolumn}
\usepackage{bm}
\usepackage{SIunits}

\begin{document}
\title{Pressure effect on superconductivity of  $A_{x}$Fe$_2$Se$_2$ ($A$ = K and Cs) }
\author{}
\author{J. J. Ying, X. F. Wang, X. G. Luo, Z. Y. Li, Y. J. Yan, M. Zhang, A. F. Wang, P. Cheng, G. J. Ye, Z. J. Xiang, R. H. Liu and X. H. Chen}
\altaffiliation{Corresponding author} \email{chenxh@ustc.edu.cn}
\affiliation{Hefei National Laboratory for Physical Science at
Microscale and Department of Physics, University of Science and
Technology of China, Hefei, Anhui 230026, People's Republic of
China\\}

\begin{abstract}
We performed the high hydrostatic pressure resistivity measurements
(up to 1.7 GPa) on the newly discovered superconductors
$A_{x}$Fe$_2$Se$_2$ ($A$ = K and Cs) single crystals. Two batches of
single crystals $K_xFe_2Se_2$ with different transition temperatures
($T_c$) were used to study the effect of pressure. The $T_c$ of the
first one gradually decreases with increasing pressure from 32.6 K
at ambient pressure. While a dome-like behavior was observed for the
crystal with $T_c=31.1$ K, and $T_c$ reaches its maximum value of
32.7 K at the pressure of 0.48 GPa. It indicates that there exists a
optimal doping with maximum $T_c$ of 32.7K in $K_xFe_2Se_2$ system.
The behavior of $T_c$ vs. pressure for $Cs_xFe_2Se_2$ also shows a
dome-like behavior, and $T_c$ reaches its maximum value of 31.1 K at
the pressure of 0.82 GPa. The hump observed in temperature
dependence of resistivity for all the samples tends to shift to high
temperature with increasing pressure. The resistivity hump could
arise from the vacancy of Fe or Se.
\end{abstract}

\maketitle

The newly discovered iron-based superconductors have attracted much
attention in past three years\cite{Kamihara, chenxh, ZARen, liurh,
rotter}. Up to now, various Fe-based superconductors, such as
ZrCuSiAs-type $Ln$FeAsO ($Ln$ is rare earth
elements)~\cite{Kamihara,chenxh,ZARen}, ThCr$_2$Si$_2$-type
$Ae$Fe$_2$As$_2$ ($Ae$ is alkali earth elements)~\cite{rotter},
Fe$_2$As-type $A$FeAs ($A$ is Li or Na)~\cite{CQJin, CWChu, Clarke}
and anti-PbO-type Fe(Se,Te)\cite{MKWu}, have been reported. The
$T_c$ of anti-PbO-type FeSe could reach 37 K under 4.5 GPa from
$T_c$$\sim$8K at the ambient pressure\cite{Medvedev}. Very recently,
by intercalating K, Rb, Cs and Tl into between the FeSe layers,
superconductivity has been enhanced to around 30 K without any
external pressure in Fe-Se system\cite{xlchen,Mizuguchi,Wang, Ying,
Krzton, Fang}, it provide a new type of iron-based superconductor to
explore high $T_c$. For the iron-pnictide, the pressure tends to
destroy the magnetic transition in the undoped compounds, and $T_c$
increases with increasing pressure for underdoped iron-pnictides,
and remains approximately constant for optimal doping, and decreases
linearly in the overdoped range\cite{Chu, Takabayashi}.
Superconductivity can be induced by pressure in the parent compounds
$AFe_2As_2$ (A=Ca, Sr, Ba, Eu)\cite{Park, Alireza, Kurita}.
Magnetism and superconductivity are strongly correlated with each
other in the iron-based superconductors. The strong pressure effect
in FeSe may be related to its strongly enhanced antiferromagnetic
spin fluctuations under pressure\cite{Imai}. Therefore, we wonder
weather the strong pressure effect still exist in $A_xFe_2Se_2$. It
is very meaningful to perform high pressure measurement in this
newly found superconductors.

In this paper, we systematically measured resistivity under the high
hydrostatic pressure up to 1.7 GPa  for the newly discovered
superconductors $K_xFe_2Se_2$ and $Cs_xFe_2Se_2$. It is found that
the transition temperature slightly increases below 0.82 GPa, and
gradually decreases with further increasing the pressure for
$Cs_xFe_2Se_2$. Two $K_xFe_2Se_2$ single crystals with different
$T_c$ were measured. For the $K_xFe_2Se_2$ crystal with
$T_c^{onset}$=32.6 K and broad hump centered at 245 K, and $T_c$
gradually decreases with increasing pressure. While $T_c$ vs.
pressure shows the similar behavior to $Cs_xFe_2Se_2$, and
$T_c^{onset}$ reaches its maximum value 32.7 K under the pressure of
0.48 Gpa for the $K_xFe_2Se_2$ with $T_c^{onset}$=31.1 K and broad
hump centered at 130 K. All these results suggest that there exists
an optimal doping with maximum $T_c$. $T_c$ increases with
increasing pressure for underdoped sample, and monotonically
decreases in the overdoped range. Resistivity hump shifts to high
temperature with increasing the pressure.

Single crystals $A_xFe_2Se_2$(A=K,Cs) was grown by self-flux method
as described elsewhere.\cite{Ying} Many shinning plate-like single
crystals can be cleaved from the final products. $Cs_xFe_2Se_2$
shows superconductivity at about 30 K, the actual compositions
determined by EDX is $Cs_{0.86}Fe_{1.66}Se_2$. Two different types
of $K_xFe_2Se_2$ were obtained. The actually composition of the
first one with the $T_c^{onset}$=32.6K is
$K_{0.85}Fe_2Se_{1.80}$(denoted $K_xFe_2Se_2$-1). The actually
composition of the second one with the $T_c^{onset}$=31.1K is
$K_{0.86}Fe_2Se_{1.82}$(denoted $K_xFe_2Se_2$-2). Pressure was
generated in a Teflon cup filled with Daphne Oil 7373 which was
inserted into a Be-Cu pressure cell. The pressure was determined at
low temperature by monitoring the shift in the superconducting
transition temperature of pure tin. The measurement of resistivity
was performed using the {\sl Quantum Design} PPMS-9.

Fig.1(a), (b) and (c) shows the temperature dependence of the
in-plane resistivity for single crystals $K_xFe_2Se_2$-1,
$K_xFe_2Se_2$-2 and $Cs_xFe_2Se_2$ under different pressures.
$K_xFe_2Se_2$-1 shows the semiconducting behavior at the high
temperature and displays a broad hump at about 245 K under ambient
pressure, and shows superconductivity at 32.6 K. The resistivity
gradually decreases and the hump becomes much more unobvious with
increasing the pressure. For $K_xFe_2Se_2$-2, the broad hump occurs
at 130 K and $T_c^{onset}$ is 31.1 K. For $Cs_xFe_2Se_2$, similar
resistivity behavior was observed with the broad hump centered at
around 285K. With  increasing the pressure, resistivity remarkably
decreases and  the hump feature becomes much more obscured.

\begin{figure}[t]
\includegraphics[width = 0.45\textwidth]{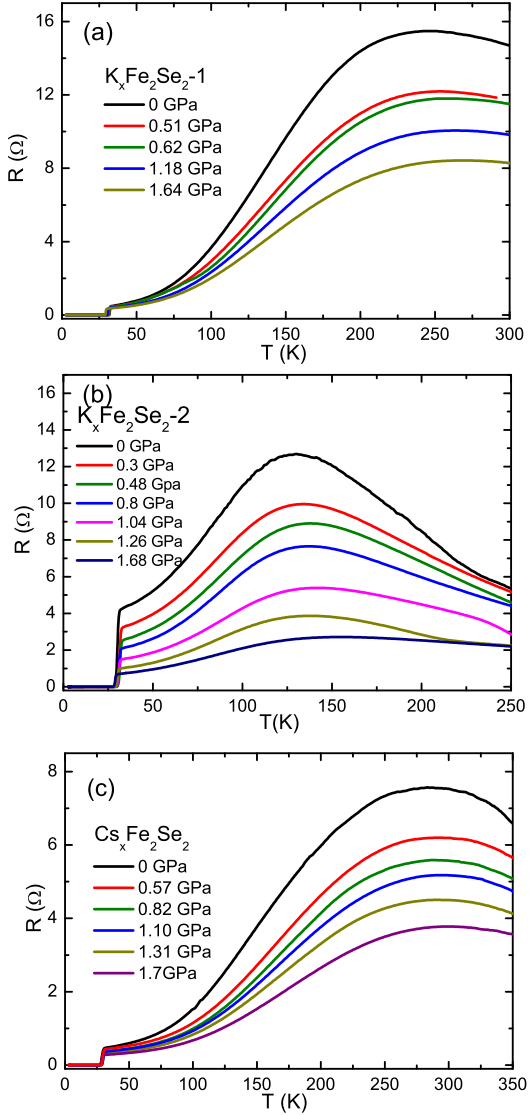}
\caption{(Color online) Temperature dependence of the in-plane
resistivity under different pressures for the single crystals: (a).
$K_xFe2Se2$-1; (b), $K_xFe2Se2$-2; and (c), $Cs_xFe2Se2$.}
\end{figure}

Fig.2(a) shows the temperature dependence of the resistivity of
$Cs_xFe_2Se_2$ under the different pressures in the low temperature
range. It indicates that the resistivity gradually decreases with
increasing pressure. We defined the $T_c$ with the temperature at
which the resistivity drops 90\%, 50\% and 10\% relative to the
resistivity just above the superconducting transition. Fig.2(b)
shows the pressure dependence of $T_c$. $T_c$ increases with
increasing the pressure below 0.82 GPa, and $T_c$ gradually
decreases with further increases the pressure. $T_c^{onset}$
increases to 31.1K under the pressure of 0.82 GPa from the 30K at
ambient pressure. $dT_c$/dP$\sim$1.3K/GPa in the region of P$<$0.82
GPa is much less than that in FeSe, and almost the same as that
observed in the electron-doped LaOFeAs\cite{Lu}. The pressure
dependence of $T_c$ in $Cs_xFe_2Se_2$ is quite similar to that of
$LaO_{1-x}F_xFeAs$ system\cite{Takahashi}.

\begin{figure}[t]
\includegraphics[width = 0.45\textwidth]{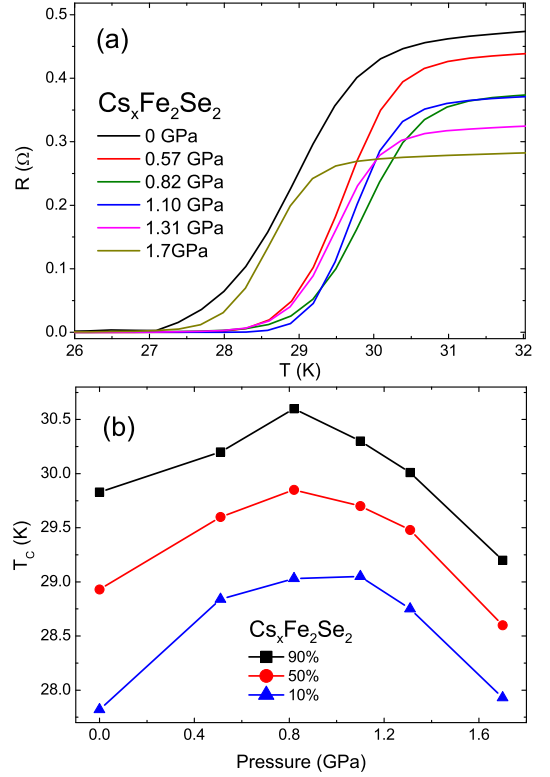}
\caption{(Color online) (a): Temperature dependence of resistivity
for single crystal $Cs_xFe_2Se_2$ under the different pressures
around superconducting transition  temperature range; (b): Pressure
dependence of $T_c$ for single crystal $Cs_xFe_2Se_2$.}
\end{figure}

Fig.3(a) shows the temperature dependence of the resistivity of
$K_xFe_2Se_2$-1 under the different pressures in the low temperature
range. The resistivity gradually reduced with increasing pressure in
the normal state. Fig.3(b) shows the pressure dependence of $T_c$.
$T_c$ monotonically decreases with increasing pressure, being
different from that shown in Fig.2b for the $Cs_xFe_2Se_2$.
$T_c^{onset}$ decreases to 29.8 K at the pressure of 1.64 GPa, which
is about 2.8 K lower than the $T_c$ at ambient pressure. The
behavior of $T_c$ vs. pressure for this compound is consistent with
the previous report\cite{guo}, while quite different from the report
by Kawasaki et al.\cite{Kawasaki} Kawasaki et al. reported that
$T_c^{onset}$ increases with increasing pressure, while $T_c^{zero}$
decreases. It suggests that the superconducting transition becomes
broadening with increasing pressure. However, the superconducting
transition width nearly does not change by pressure as shown in
Fig.3(a). Such difference could arise from the quality of single
crystal or inhomogeneity of applied pressure.

\begin{figure}[t]
\includegraphics[width = 0.45\textwidth]{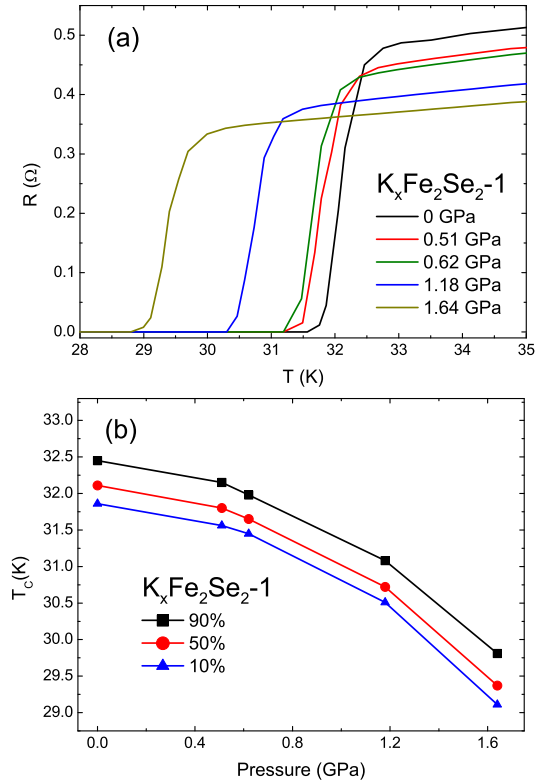}
\caption{(Color online) (a): Temperature dependence of resistivity
for single crystal $K_xFe_2Se_2$-1 under the different pressures
around superconducting transition  temperature range; (b): Pressure
dependence of $T_c$ for single crystal $K_xFe_2Se_2$-1.}
\end{figure}

Fig.4(a) shows the temperature dependence of the resistivity of
$K_xFe_2Se_2$-2 under the different pressure around the temperature
range of superconducting transition. The $T_c$ at ambient pressure
is 1.5 K lower than that of $K_xFe_2Se_2$-1. The behavior of $T_c$
vs. pressure is quite different from that of $K_xFe_2Se_2$-1 as
shown in Fig.3(b). $T_c$ as a function of pressure shows a dome-like
behavior as shown in Fig.4(b). It is similar to that observed in
single crystal $Cs_xFe_2Se_2$. $T_c^{onset}$ gradually increases
with increasing pressure and reaches the maximum value of 32.7 K at
the pressure of 0.48 GPa, then monotonically decreases with further
increasing pressure. It should be addressed that the $K_xFe_2Se_2$-1
and $K_xFe_2Se_2$-2 shows different pressure dependence of $T_c$.
Such different pressure dependence of $T_c$ indicates that there
exists optimal doping with the maximum $T_c\sim 32.7$ K in
$K_xFe_2Se_2$ system.  The pressure tends to destroy the magnetic
transition in the undoped compounds and $T_c$ increases with
increasing pressure for underdoped iron-pnictides, and monotonically
decreases in the overdoped range. The different effect of pressure
on $T_c$ between the crystals $K_xFe_2Se_2$-1 and $K_xFe_2Se_2$-2 is
easily understood because the $K_xFe_2Se_2$-1 is in the slightly
overdoped range, while the $K_xFe_2Se_2$-2 is in the underdoped
range.

\begin{figure}[t]
\includegraphics[width = 0.45\textwidth]{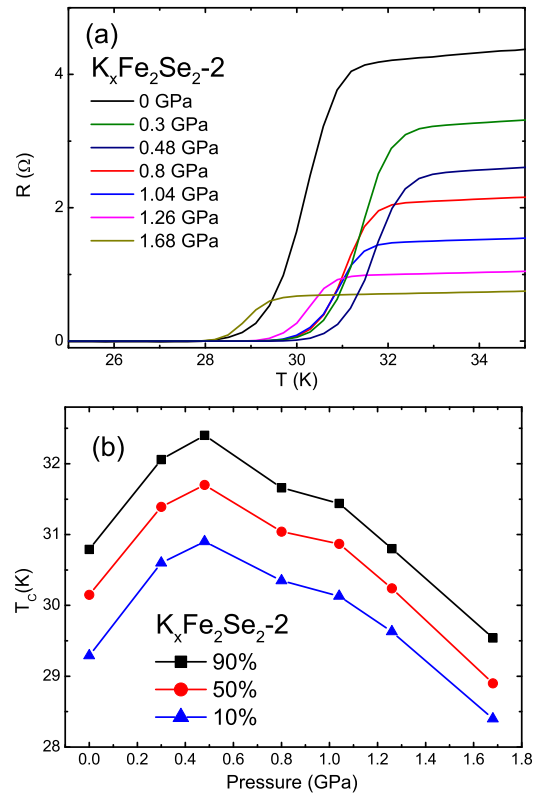}
\caption{(Color online)  (a): Temperature dependence of resistivity
for single crystal $K_xFe_2Se_2$-2 under the different pressures
around superconducting transition  temperature range; (b): Pressure
dependence of $T_c$ for single crystal $K_xFe_2Se_2$-2.}
\end{figure}

Fig.5 shows the pressure dependence of the resistivity hump
temperature for $Cs_xFe_2Se_2$, $K_xFe_2Se_2$-1 and $K_xFe_2Se_2$-2.
We defined the hump temperature when the resistivity reaches its
maximum value. The temperature of the hump monotonically increases
with increasing pressure for all the samples. The shift of the
resistivity hump temperature for $Cs_xFe_2Se_2$ is very small below
0.82 GPa and increases apparently in the high pressure range. The
temperature of the hump shows no direct connection with the $T_c$
because the pressure dependence of hump temperature is quite
different from that of $T_c$. The origin of the hump in this kind of
superconductors could arise from the content of K or the vacancy of
Fe or Se sites.

\begin{figure}[t]
\includegraphics[width = 0.45\textwidth]{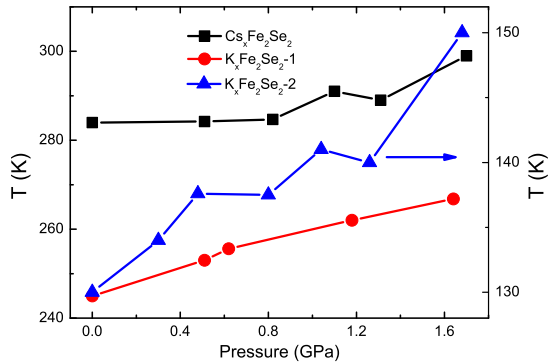}
\caption{(Color online) Pressure dependence of the resistivity hump
temperature for single crystals $Cs_xFe_2Se_2$, $K_xFe_2Se_2$-1 and
$K_xFe_2Se_2$-2. }
\end{figure}

The normal state resistivity behavior and pressure dependence of
$T_c$ are quite different for the crystals $K_xFe_2Se_2$ with
different $T_c$. The $T_c$ as a function of pressure for
$K_xFe_2Se_2$-2 shows a dome-like behavior and reaches its maximum
$T_c$ at the pressure of 0.48 GPa. While $T_c$ is gradually
suppressed with increasing the pressure for $K_xFe_2Se_2$-1. The
pressure dependence of $T_c$ for $K_xFe_2Se_2$-2 and $Cs_xFe_2Se_2$
is quite similar to that observed in the $LaO_{1-x}F_xFeAs$ and
parent iron-pnictides\cite{Takahashi, Takahashi1, Kurita}. While for
$K_xFe_2Se_2$-1, the pressure dependence of $T_c$ is similar to that
observed in the overdoped iron-pnictides\cite{Gooch}. The maximum
$T_c$ for the two batched of single crystals $K_xFe_2Se_2$ is the
same, and reaches at ambient pressure and at 0.48 Gpa, respectively.
The difference of $T_c$ between the two batches of single crystals
$K_xFe_2Se_2$ is very small (just 1.5 K), while the temperature of
the resistivity hump is 245 K and 130 K, respectively. It indicates
the hump temperature strongly depends on the sample. The hump
behavior could originate from the vacancy of Fe or Se. Another
evidence is that the normal state resistivity is very high compared
with other iron-pnictide superconductors because the vacancy in
conducting FeSe layers leads to a strong scattering, consequently
high resistivity. It suggests that the physical behavior of
$A_xFe_2Se_2$ is very sensitive to the deficiency, and change of the
deficiency strongly affects the normal state resistivity although
the $T_c$ does not change too much.

In conclusion, we performed the high hydrostatic pressure
resistivity measurement for the newly discovered superconductors
$A_xFe_2Se_2$(A=K, Cs). For $Cs_xFe_2Se_2$, $T_c$ starts to increase
at the pressure less than 0.82 GPa, and $T_c$ decreases with further
increasing the pressure. This behavior is similar to that in
$K_xFe_2Se_2$-2. While the behavior is quite different for
$K_xFe_2Se_2$-1 with $T_C^{onset}$=32.6K, $T_c$ monotonically
decreases with increasing pressure. The different pressure
dependence of $T_c$ between these single crystals is because they
are in different doping level with different $T_c$. The temperature
of resistivity hump increases with increasing the pressure. The
resistivity hump could arise from the deficiency of Fe or Se in
conducting layers.

{\bf ACKNOWLEDGEMENT} This work is supported by the Natural Science
Foundation of China and by the Ministry of Science
and Technology of China, and by Chinese Academy of Sciences.\\

\end{document}